# FERROELECTRIC SOFT MODES IN CERAMICS AND FILMS

Jan Petzelt, Stanislav Kamba and Jiri Hlinka

Institute of Physics, Academy of Sciences of the Czech Republic
Na Slovance 2, 182 21 Praha 8, Czech Republic

## ABSTRACT

Ferroelectric soft mode properties in ceramics and thin films are reviewed and compared with single crystals. In general, the permittivity in ceramics and thin films is markedly reduced compared to the single crystals mainly because of low-permittivity grain boundaries - dead layers - (sometimes also nano-cracks) present in the samples. In $SrTiO_3$ and $BaTiO_3$, this dielectric reduction is fully explained by stiffening of the soft mode frequency. In most other ferroelectrics, the soft mode accounts only partly (or not at all) for such a permittivity reduction, since the transition is only partially driven by the phonon mode softening. This can be assigned to the crossover from displacive behaviour (far from $T_C$ dominated by the soft phonon contribution to the dielectric response) to the order-disorder regime (close to $T_C$ dominated by the central mode contribution). In these cases, additional relaxational dispersion below the optical phonon range (central mode) accounts for the main part of the dielectric anomaly near $T_C$. Influence of nano-cracks on reduced permittivity near $T_C$ and related stiffening of the central mode is manifested in the antiferroelectric $PbZrO_3$ ceramics. In perovskite relaxor ferroelectrics, the smeared and frequency dependent permittivity maximum is caused by a strong dielectric relaxation assigned to the dynamics of polar clusters. This relaxation slows down and its frequency spectrum anomalously broadens on cooling. Like in the normal ferroelectrics, permittivity in relaxor ceramics and thin films reduces with decreasing grain size and/or film thickness. It can be also explained by the low-permittivity grain boundaries affecting the relaxational dynamics. The (infrared active) soft phonon mode of relaxors does not show any anomaly near the permittivity maximum $T_{max}$, but it partially softens towards the Burns temperature. In relaxor–like ferroelectrics, the ferroelectric transition (as a rule of first order) is neither displacive, nor classically order-disorder, nor does it reveal a crossover between both types of dynamic behaviour. At ferroelectric transitions in these materials, an abrupt size increase of the polar clusters, changing below $T_C$ into ferroelectric domains, seems to account for the dynamic manifestation of the ferroelectric transition, without appreciable phonon softening.

# 1. SOFT MODES AND CRITICAL DYNAMICS NEAR FERROELECTRIC PHASE TRANSITIONS IN CRYSTALS

Soft phonon modes in crystals are low-frequency lattice vibrational modes whose frequencies tend to zero (soften) if the crystal approaches a structural phase transition on changing the external thermodynamic force (temperature, pressure, field). At a (second order) phase transition, the crystal structure looses its stability against the vibration given by the eigenvector of the soft mode. Such a picture holds for well-ordered and relatively weakly anharmonic crystal lattices and describes so called displacive phase transitions. In this case the classical assumptions of the thermodynamic Landau theory of phase transitions [1] imply [2] that the soft-mode frequency $\omega_{SM}$ versus temperature should obey the famous Cochran law [2, 3, 4]

$$\omega_{SM}^2 = A(T - T_C), \qquad (1)$$

where $T_C$ is the critical Curie temperature and $A$ is a constant. For proper ferroelectric transitions, where the order parameter is polarization, the soft mode is infrared (IR) active in both paraelectric and ferroelectric phases and we may readily assume that the IR oscillator strength $S_{SM} = \Delta\varepsilon_{SM}\omega_{SM}^2$ of the soft mode is temperature independent across the phase transition [5]. Cochran law then implies the Curie-Weiss law for the static permittivity:

$$\varepsilon_0 = \varepsilon_B + C/(T - T_C), \qquad (2)$$

where the Curie-Weiss constant $C = S_{SM}/A$ is determined by the temperature dependence of soft mode frequency and $\varepsilon_B$ is the sum of electronic contribution and contributions from other polar transverse optical (TO) modes which can be considered as temperature independent.

Frequently, the materials undergoing ferroelectric transitions are partly structurally disordered and/or strongly anharmonic and the simple soft-mode picture does not hold. Let us consider that the ions in one sublattice may assume two (or more) close sites in the lattice of the disordered (high-temperature) phase. Their dynamical hopping among available sites gives rise to a dielectric relaxation (in the simplest case Debye-like) below the polar phonon frequencies (typically in the microwave (MW) range), which slows down when the crystal approaches the phase transition. This is so-called critical slowing down and the relaxation frequency $\omega_R$ (given by the maximum in the dielectric loss vs. frequency spectra) in the classical form obeys

$$\omega_R = A_R(T - T_C), \qquad (3)$$

which again in the static limit yields the Curie-Weiss law for the permittivity. This behaviour is typical for order-disorder transition. It should be noted that in the displacive case, if the soft phonon damping $\Gamma$ is essential, the soft mode becomes overdamped ($\Gamma > \omega_{SM}$) and it may become very difficult to determine experimentally the soft mode frequency $\omega_{SM}$. In fact, the response of a heavily overdamped oscillator approaches that of the Debye relaxation in the frequency range of the dielectric-loss maximum and below it [5]. In this case the maximum in the dielectric loss spectra is a more representative and easily experimentally determined frequency which approaches the relaxation frequency and linear critical slowing down with $\omega_R = \omega_{SM}^2/\Gamma$.

In many cases the dynamic anomalies around a ferroelectric phase transition are more complicated and contain features of both displacive as well as order-disorder behaviour. Typically, far away from the transition some phonon softening appears which ceases close to the transition point where additional relaxation appears and usually contributes substantially to the permittivity maximum. Such a relaxation is frequently called central mode in analogy to inelastic scattering experiments [6], where such excitations contribute as quasielastic peaks with the half-width equal to $\omega_R$. Such a situation is usually referred to as crossover from displacive to order-disorder behaviour.

Let us note that the laws (1-3) assume the situation where the fluctuations of the order parameter can be neglected, as in the usual Landau theory. These laws will be valid also in a close vicinity of a second-order transition point in case of system with mean-field critical behaviour. Such a mean-field critical behaviour is indeed expected to be fulfilled for the proper uniaxial ferroelectrics. For ferroelectrics with a multi-component order parameter, for weak [7] and improper ferroelectrics and for non ferroelectric structural transitions, deviations from the mean-field-like laws (1-3) are expected [8], in the (usually narrow) critical region.

So far we have dealt with well-defined sharp phase transitions appearing in macroscopically and nanoscopically homogeneous systems. However, many systems of great practical importance (doped crystals, solid solutions, ceramics, thin films, composites etc.) are not perfectly homogeneous and may undergo smeared diffuse phase transitions or even relaxor or dipolar glass behaviour without changes in the macroscopic symmetry. In such systems there is no such a simple way to study the critical behaviour [9], which might be connected with freezing or glass transitions. Nevertheless, the soft and central mode concept may be used even here to describe the usually observed pronounced temperature changes in the dynamic behaviour [10].

## 2. EFFECTIVE DIELECTRIC AND INFRARED RESPONSE IN INHOMOGENEOUS DIELECTRICS

If the sample to be dielectrically characterized by a weak external probing field is inhomogeneous, one can usually determine only its effective dielectric response, which does not give full information about the spatial variation of the dielectric function. It is often useful to consider the inhomogeneous material as a densely filled micro-composite formed by individual, dielectrically homogeneous parts. Depolarization field, which always appears on the boundaries between the different parts, tends to reduce the probing electric field in higher-permittivity parts (condition of the constant normal part of the electric displacement ***D*** through the boundary). As the result, there can be pronounced differences between the effective dielectric response and the response of individual homogeneous parts.

The magnitude of the depolarization field depends on the topology and shape of individual parts. Let us consider a composite medium in ideal plane-capacitor geometry. If the boundaries between the parts are only parallel to the probing field, there is no depolarization field effect and the effective dielectric response is simply the arithmetic average of individual responses weighted by their volume concentrations (equivalent circuit of parallel capacitances). On the other hand, the maximum depolarization field effect belongs to the planar geometry of individual parts with the probing electric field normal to the planar boundaries (equivalent circuit of series capacitances). The situation in real systems should always appear somewhere in between. The higher is the depolarization field, the higher is the difference between simple averaging the permittivity of individual parts and effective dielectric response, making the influence of low-permittivity parts more important. If the composite geometry is close to a random mixture of spherical individual parts, a simple physical insight into the observed effective properties in terms of dielectric response of individual homogeneous parts provides the famous Bruggeman's effective medium approximation (EMA) [11, 12].

The general approach to the two-component composites with arbitrary geometry was developed by Bergman [12]. For two components with isotropic complex permittivities $\varepsilon_1$ and $\varepsilon_2$ and volume concentrations $x_1$ and $x_2 = 1-x_1$, the effective dielectric response can be written in the form [13]:

$$\varepsilon_{eff} = V_1(x_1)\varepsilon_1 + V_2(x_2)\varepsilon_2 + \int_0^1 \frac{x_1}{(1-n)} G(n, x_1) \frac{\varepsilon_1 \varepsilon_2}{(1-n)\varepsilon_2 + n\varepsilon_1} dn, \tag{4}$$

where $n \in (0, 1)$ is a generalized depolarization factor and $G$ is called spectral density function, fulfilling the normalization condition

$$V_1 + V_2 + \int_0^1 \frac{x_1}{1-n} G(n, x_1) dn = 1. \tag{5}$$

The first two terms in Eq. (4) describe responses which are not influenced by the depolarization field, just weighted by factors $0 \leq V_1 \leq x_1$ and $0 \leq V_2 \leq x_2$. The physical meaning of $V_1$ and $V_2$ is the volume part of component 1 and 2, respectively, percolated in the direction of probing field, since only such part of the sample volume is not influenced by the depolarization field. The third term in Eq. (4) describes the remaining volume part which is influenced in various manners by the depolarization field. The function $G(n)$ characterizes the geometry, topology and interaction of individual homogeneous regions. The concrete solvable effective-medium models or empirical mixing formulas like Maxwell-Garnett, Bruggeman, Looyenga, Lichtenecker, Hashin-Shtrickman etc. correspond to a particular choice of the spectral density function $G$ and percolated volumes $V_1$ and $V_2$ [13, 14, 15].

The above formulation is particularly useful in discussing the response of granular materials (ceramics, polycrystalline films) and core-shell composites with 0–3 connectivity, in which the grain boundary region (shell) plays the role of the second component whose dielectric properties may differ from that of the bulk (core). In high-permittivity materials this region has usually lower permittivity (in ferroelectrics it might be non-ferroelectric - dead or passive layer) and in semiconductive ceramics it may differ in conductivity (e. g. blocking boundaries in grain-boundary capacitors). In this geometry the bulk properties are not percolated at all (each core is surrounded by the shell, $V_1 = 0$), and if the total shell volume is small, $x_2 << x_1$, and the function $G(n)$ is nonzero only for one value of $n$ [16], then

$$\varepsilon_{eff} = \left(1 - \frac{x_1}{1-n}\right)\varepsilon_2 + \frac{x_1}{1-n} \frac{\varepsilon_1 \varepsilon_2}{(1-n)\varepsilon_2 + n\varepsilon_1} \tag{6}$$

This is so called generalized brick-wall model [16], which is valid for bricks of arbitrary (but similar) shapes and sizes covered with thin walls of any thicknesses, characterized by just a single parameter $n$, $0 \leq n \leq x_2$. For $n = (1/3)x_2$ the frequently used standard brick-wall model with cubic bricks and coated-spheres model (Hashin-Shtrickman) are recovered [16] (both models are equivalent in the limit of small $x_2$).

Another important case, where effective medium approaches are essential, concern polycrystalline ceramics, films or other one-component composites in which the individual crystal grains are anisotropic. Stroud [17] has shown that at the level of EMA approximation (spheroidal particles), such systems are equivalent to a composite of isotropic particles whose dielectric responses are equal to principle dielectric responses of the anisotropic crystallite. For instance, a dense and macroscopically isotropic polycrystal or ceramics comprised of dielectrically uniaxial elliptical crystallites is equivalent to a two-component composite of isotropic particles with the corresponding two principal dielectric responses and with the volume concentration of 1/3 and 2/3 for the extraordinary and ordinary response, respectively. Such approach was successfully used for fitting the IR reflectivity of non-ferroelectric optically uniaxial pressed pellets with various particle shapes [18].

Particular advantage of separating the percolated and non-percolated part of the effective dielectric response in Eqs. (4) and (6) becomes evident when discussing the effective ac response. Bergman theory and mixing formulas like EMA are equally well valid for ac response, as long as the *E*-homogeneity condition is fulfilled. In dielectrics it is valid up to the IR range including polar phonon absorption as long as the probing radiation wavelength and penetration depth of the radiation are much larger than the particle size. Generally, the ac dielectric response consists of several relaxational and/or resonance dispersion regions in the spectrum, which are characterized by poles of the complex permittivity in the complex frequency plane. Resonances are mostly due to polar TO phonon mode absorption in the IR (and piezoelectric resonances if available), relaxations at lower frequencies appear in impure, inhomogeneous, and/or strongly anharmonic systems. In the percolated part of the response (first two terms in Eq. (4)), the frequencies of these poles remain unchanged so that all dispersion regions in the ac response remain un-shifted, appear just weaker by the factor of $V_1$ or $V_2$. The non-percolated part of the response is, however, influenced by the depolarization field in the way that all dispersion regions may be shifted up in the frequency, which e.g. may result in the reduced effective static response [14, 19]. These shifts, which result in stiffening of the TO frequencies in the latter case, are the greater the stronger are the dielectric strengths $\Delta\varepsilon$ (contributions to the static permittivity) of the dispersion in question. The renormalized modes are sometimes called geometrical resonances (Fröhlich modes or surface modes in case of isolated particles). If the assumption of single $n$ in Eq. 6 is released (more complex topology and shape of the particles or grains), an absorption continuum is obtained instead of each single geometrical resonance and one can expect smearing of the shifted-up TO modes as easily seen from the general Bergman Eq. 4.

## 3. FERROELECTRIC SOFT MODE CONCEPT IN POLYCRYSTALLINE CERAMICS, POWDERS AND FILMS

In a ferroelectric composite with nonferroelectric (insulating) boundaries of smaller and essentially temperature independent permittivity (dead or passive layers) soft and central modes are strongly shifted up and their softening near $T_C$ is effectively blocked, as their dielectric strength $\Delta\varepsilon$ is by far the highest in the dielectric spectrum. For instance, in the paraelectric phase of a ferroelectric ceramic, softening of the ferroelectric soft mode or slowing down of the critical relaxation is blocked so that the Curie-Weiss law should be modified, i. e. Curie temperature should be shifted down and the permittivity maximum at the ferroelectric transition is reduced. This is exactly what was observed in $BaTiO_3$ nanoceramics [20, 21, 22]. Obviously, powder samples can be treated as a composite of the non-percolated bulk-material particles and the matrix in which the particles are embedded. Therefore, a pronounced influence of the depolarization field and stiffening of strong polar mode frequencies (particularly no mode softening in ferroelectrics) is expected here, too [23, 24].

Properties of the soft modes in thin films should obviously reflect the substrate-induced stresses, but in addition to that, influence of grain boundary effects and interfacial layers between the film and substrate or electrode can be very important or even dominant. These interfacial nano-layers, which may show different dielectric properties than the film bulk, may seriously influence the effective dielectric response perpendicular to the film plane (usual plane capacitor geometry for dielectric measurements), but are of minor influence for the in-plane response (as probed by the IR wave when using normal incidence), again due to the difference in the depolarization field effect, which is maximized in the former case and zero in the latter one.

Finally, let us note that usually the transition temperature $T_C$ might not be appreciably influenced even in the case of suppressed dielectric anomaly at the ferroelectric transition, since the effect of depolarization field can be compensated by diffusing charges which, however, usually (at not too high temperatures) cannot follow the frequencies of dielectric measurements (>100 Hz). So, in such cases the Curie-Weiss divergence of the permittivity may be expected only for very low frequencies below the reciprocal time of depolarization field compensation.

## 4. SOFT MODE BEHAVIOUR AND LOCAL SYMMETRY IN PEROVSKITE RELAXOR FERROELECTRICS

Prototype relaxor ferroelectric materials like $PbMg_{1/3}Nb_{2/3}O_3$ (PMN) do not undergo classical macroscopic structural phase transition so that no typical soft mode behaviour is expected. Relaxor materials are nevertheless characterized by appearance of polar nano-regions (of typical size of the order of 20-200 Å)[25, 26], usually at a relatively high temperature called Burns temperature $T_B$ (typically 600-700 K). These clusters are assumed to undergo very fast dynamic motion close to $T_B$ (in the $10^{11}$-$10^{12}$ Hz range) which gradually slows down on cooling and finally freezes at some freezing temperature $T_f$ (typically around 200 K) – see Fig. 1. The strong dielectric contribution of such clusters (permittivity of the order of $10^3$-$10^5$) is maximal near some temperature $T_m$ (which is, however, frequency dependent) appearing typically 50-100 K above $T_f$. The corresponding dielectric dispersion is of relaxational nature; it approaches the simple Debye relaxation form near $T_B$ but broadens substantially at lower temperatures (polydispersive behaviour). On cooling, the characteristic frequencies show slowing down, the spectrum sometimes splits into two or even three broad relaxation regions which fuse together into frequency independent loss spectrum below $T_f$ (so called 1/f noise) [28], while the permittivity undergoes logarithmic dispersion and gradually decreases on further cooling [27, 28, 29, 30]. The freezing temperature $T_f$ is usually characterised by the Vogel-Fulcher slowing down of the lowest-frequency relaxation component

$$\omega_R = \omega_{R\infty} \exp[-E_A/(T-T_f)] . \qquad (7)$$

Close to $T_f$ the mean relaxation time changes into the Arrhenius behaviour, but this cannot be followed to much lower temperatures because of the extreme relaxation broadening. Anyway, the temperature dependence of the permittivity and its dispersion up to the IR range are dominated by the contribution of polar clusters. The microscopic mechanism of this huge contribution can be so far discussed only very qualitatively [31, 32]. It seems clear that at high temperatures close to $T_B$ the flipping of the whole clusters may be the dominant mechanism, but on cooling the clusters grow and slow down and the breathing mechanisms (fluctuations in the cluster walls, or, in other words, cluster volume fluctuation) should become dominant. The latter mechanism clearly prevails below $T_f$ where the clusters should be frozen. Freezing concerns the central part of each cluster only, but their walls (probably relatively broad) are apparently still actively vibrating.

Let us ask what can be expected concerning the polar phonon behaviour and its softening in such materials. Already at very high temperatures above $T_B$, the relaxors usually show Curie-Weiss growth of the permittivity, which should be caused by softening of the lowest polar mode in displacive systems like perovskites. Formation of polar clusters should be connected with local softening of these phonons. However, since the clusters obviously appear separately from each other, they are not percolated close to $T_B$ and therefore the softening should be obscured in the effective IR response and no dielectric anomaly at $T_B$ may be expected, in agreement with experiments. Below $T_B$, the expected polar phonon behaviour is complicated because the polar regions may become quite strongly anisotropic with much smaller dielectric response along the local polarization than perpendicular to it [33]. Therefore the splitting of both the local and effective soft modes have to be considered and moreover, the relaxation (central mode) due to the cluster dynamics appears in the same spectral range ($\sim 10^{12}$ Hz) near below $T_B$ – see Fig. 1. Polar phonon spectra of relaxor ferroelectrics have been recently reviewed in [34]. In sections 7 and 8, we shall describe in more detail the low frequency dynamics of PMN and PST.

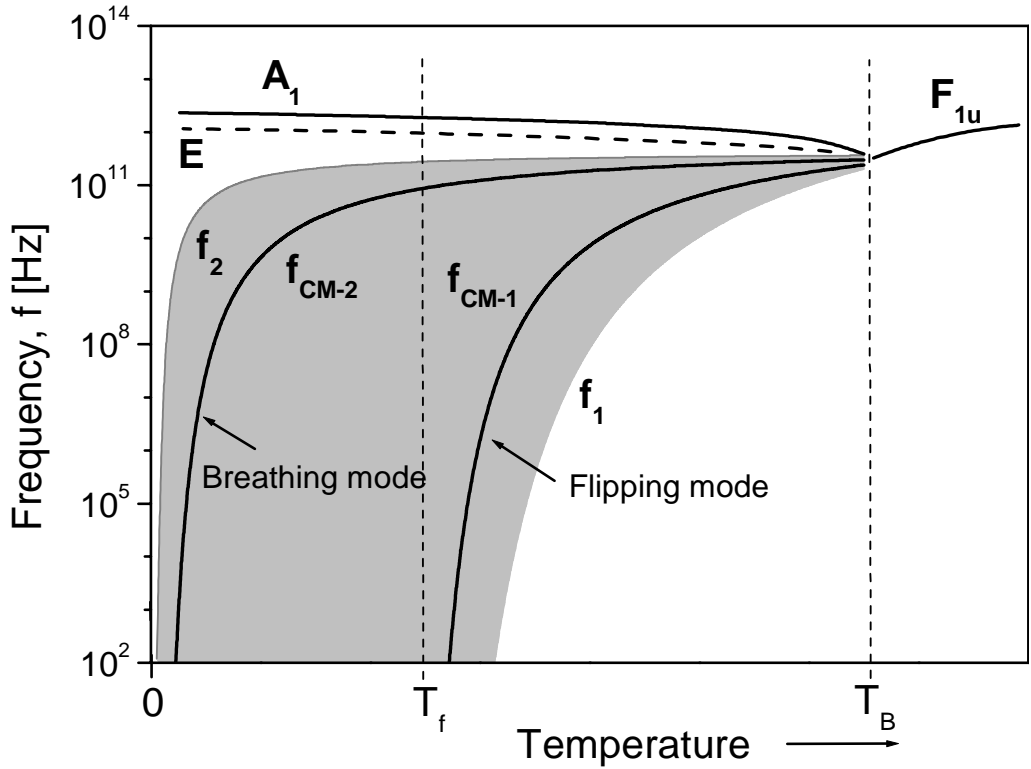

Fig. 1. Schematic picture of the temperature dependences of all anomalous excitations in perovskite relaxor ferroelectrics. Ferroelectric soft mode of $F_{1u}$ symmetry splits below $T_B$ into $A_1$ and E components, which harden on cooling. Relaxation (central) mode appears below $T_B$, broadens on cooling and splits into two components describing flipping and breathing of polar nanoclusters. Grey area (limited by $f_1(T)$ and $f_2(T)$ curves) marks the frequency and temperature range, where the distribution function of relaxation frequencies is nonzero.

## 5. IR AND RAMAN SOFT MODE SPECTROSCOPY IN PURE AND DOPED SrTiO$_3$ CERAMICS AND FILMS

SrTiO$_3$ (STO) is a classical incipient ferroelectric with strongly increasing permittivity on cooling. It is well known since the 1994 [35] that in STO thin films the permittivity is smaller and does not increase as much as in single crystals. It is thickness dependent, but even in the limit of very thick films it does not reach the crystal values [36]. The thickness dependence was assigned to low-permittivity (dead) interfacial layers between the film and electrodes [36]. In addition, strains induced by the substrate may also influence the dielectric response. However, in polycrystalline films also grain boundaries play an important role [16, 19]. To separate the latter effect from the others mentioned it is of interest to study bulk ceramics, where the grain boundary effect should dominate. This was the motivation of the dielectric and soft mode study in dense STO ceramics (of averaged grain size ~1500 nm) [37]. Later the soft mode behaviour in polycrystalline thin films with much smaller grain size was studied [38]. In IR experiments mostly the in-plane response is probed without electrodes, which minimises the effect of interfacial layers. Very recently, the work was continued by studying also the dense fine-grain ceramics with the averaged grain size of 150 nm [39, 40]. The main results are summarised as follows.

From standard dielectric measurements, a dramatic decrease (more than by an order of magnitude) in the low-temperature permittivity was revealed with the smaller grain size. IR reflectivity proved that the soft mode behaviour is fully compatible with this reduction. The temperature dependence of its frequency is shown in Fig. 2. Note the strong soft mode stiffening in fine-grain ceramics and film, particularly at low temperatures. Both sets of data on ceramics for all temperatures can be well fitted within the brick-wall model by a simple dead-layer on cubic bricks (or coated-spheres model) with the result that the ratio of the dead-layer permittivity $\varepsilon_{gb}$ and thickness $d$ (in nm) should fulfill $\varepsilon_{gb}/d \cong 10$ (nm$^{-1}$). Strong correlation between both quantities (caused by the essentially equivalent circuit of series capacitances) did not allow the authors to determine both parameters independently. From careful comparison of Raman spectra they obtained some complementary information. Appearance of forbidden IR modes in the Raman spectra indicates at least local acentricity in the structure. It was suggested that this is due to frozen polarization near grain boundaries [37], which develops in order to screen the charged O vacancies in the boundaries [41].

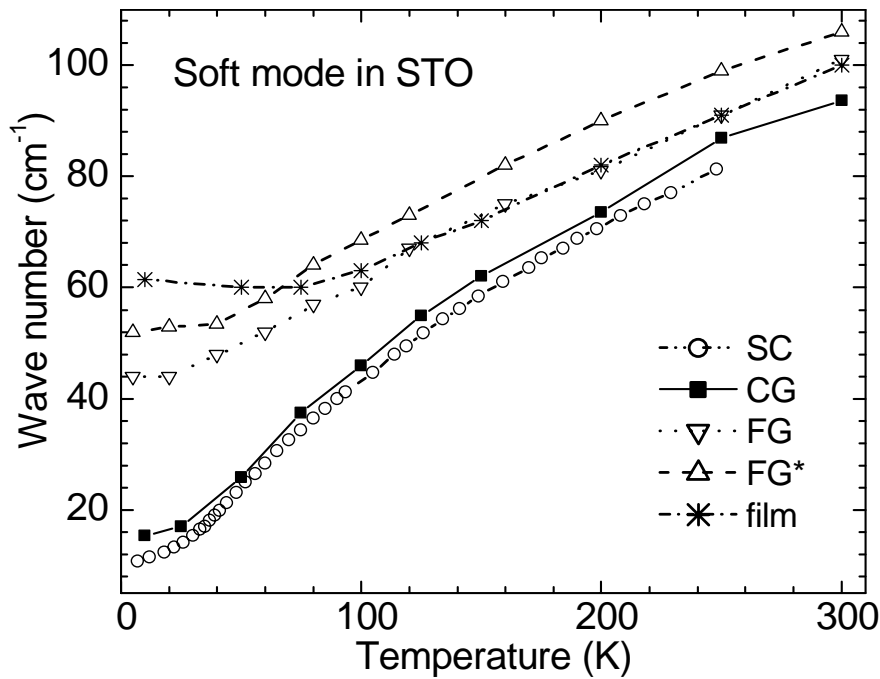

Fig. 2. Soft-mode frequencies of STO single crystal (SC), coarse-grain ceramics (CG, grain size ~1500 nm), fine-grain (~150 nm) ceramics (FG – 99% density and FG* - 98% density) and polycrystalline film (thickness 720 nm) of the same, but columnar grain size as FG ceramics. (After [40]).

Moreover, after subtracting the one-phonon response, a strong background scattering appeared in the low-temperature Raman spectra with the peak value of ~70 cm$^{-1}$. Assuming polarized perovskite lattice near grain boundaries, this should correspond to smeared soft phonon response localized in the polarized regions with averaged permittivity value of ~500. The thickness of these layers is given by the correlation length of polarization fluctuations, which at low temperatures is estimated to be less than 5 nm [40]. However, to account for the full grain size effect of the dielectric response, additional very thin grain boundary layer (~1 nm) was needed with a very small local permittivity (~8). This was estimated using a brick – double-wall model [40]. The thin grain-boundary layer for particular grain boundaries in bicrystal was studied by HRTEM [42] and *ab-initio* calculations [41] and it was shown that it is really very thin (below 1 nm) and contains defect pentagonal units with O vacancies, which probably give the reason for the frozen polarization around the grain boundaries.

Let us note that the appearance of the TO1 soft mode in the Raman spectra of fine-grain ceramics differs from that in IR spectra. The stiffening compared to single crystals is not as strong as in the IR response and the $E_u$-symmetry component of the TO1 doublet below the structural transition into the tetragonal phase (at ~107 K) couples bilinearly with the $E_g$-symmetry component of the structural soft-mode doublet. Such a coupling is allowed only in the polarized regions. It can be shown that the E modes can propagate mainly perpendicularly to the grain boundaries through the polarized regions, whereas the A modes are localized inside the non-polarized grain bulk with much higher permittivity. This enables the coupling observed in the Raman spectra and also explains the smaller stiffening. It is caused by small depolarization field effects for the TO(E) modes, where the ***E***-field is mostly along the grain boundaries (geometry of parallel capacitors). Its smaller but non-zero value is caused only by the finite wavelength of the Raman active phonon, which is comparable to the grain size so that the EMA is not strictly valid. On the other hand, in the IR response the phonon wavelength is much larger than the grain size, which justifies using the EMA and makes the effect of the depolarization field more pronounced. So, in inhomogeneous media the measured effective polar phonon-mode frequencies (particularly the strong ferroelectric soft modes) in Raman and IR response may appreciably differ due to their different wavevectors and propagation properties.

Additional discussion is needed for thin films. The TO1 mode stiffening compared to single crystals was first time observed in the IR transmission [43] and then by IR ellipsometry [44] and Raman response [45]. The reasons for it were most thoroughly studied in [38]. In addition to the grain boundary effects in polycrystalline films, a very strong effect is provided by possible nano-cracks in thicker films. In Fig. 3, the effects of cracks along the grain boundaries on permittivity and the TO1 frequency were calculated using the brick-wall model. It is seen that the observed data from 2 films differing just by factor of 2 in the thickness can be explained by 0.2 and 0.4 % of crack-type porosity. On the other hand, in epitaxial thin films the stresses from the substrate dominate. The phase diagram of STO under uniaxial (biaxial) pressure was theoretically calculated from the known thermodynamic parameters [46, 47] and it appears very sensitive to both compressive as well as tensile stresses. Using appropriate substrate, the ferroelectric transition may be induced at low temperatures and even shifted up. E.g., the tensile in-plane stress produced by $DyScO_3$ substrate may induce the in-plane ferroelectricity up to room temperature [48]. In [38] the quasi-epitaxial film on sapphire exhibited ferroelectric transition near 120 K together with TO1 softening more pronounced than in single crystals - see Fig. 4. Below the structural transition appearing probably at the same temperature, the TO1 mode couples with the structural soft mode doublet as seen in Fig. 4. On the other hand, the compressive stress induced by a $NdGaO_3$ substrate induces the out-of plane ferroelectricity up to ~150 K [49] so that in such strained films the in-plane and out-of-plane ferroelectric instabilities should be considered separately. In the latter case, the in-plane TO1 mode was stiffened up to 132 cm$^{-1}$ at room temperature and softens at low temperatures only down to 108 cm$^{-1}$, but the out-of-plane ferroelectric transition was indicated by the appearance of the silent TO3 mode near 318 cm$^{-1}$ in the IR spectra.

An important issue for applications concerns the *E*-field tunability of the permittivity. Since the high permittivity is fully due to the TO1 mode contribution, it is expected that the tunability is caused by stiffening of this mode in the bias field. This effect was first studied by

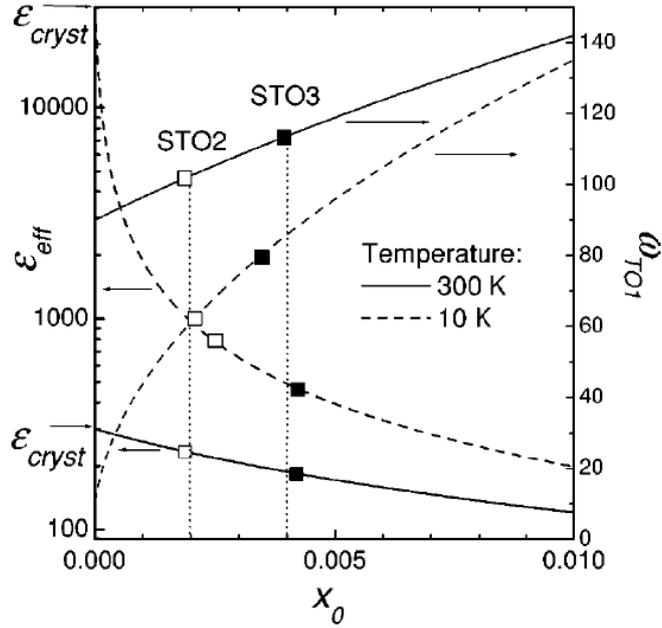

Fig. 3. Effective permittivity and TO1 frequency as a function of crack-type porosity $x_0$ between some columnar grains in STO films from the brick-wall model. The experimental data on films correspond to 0.2% and 0.4% of percolated porosity for the STO2 (360 nm thick) and STO3 (720 nm thick) sol-gel films, respectively. (After [38]).

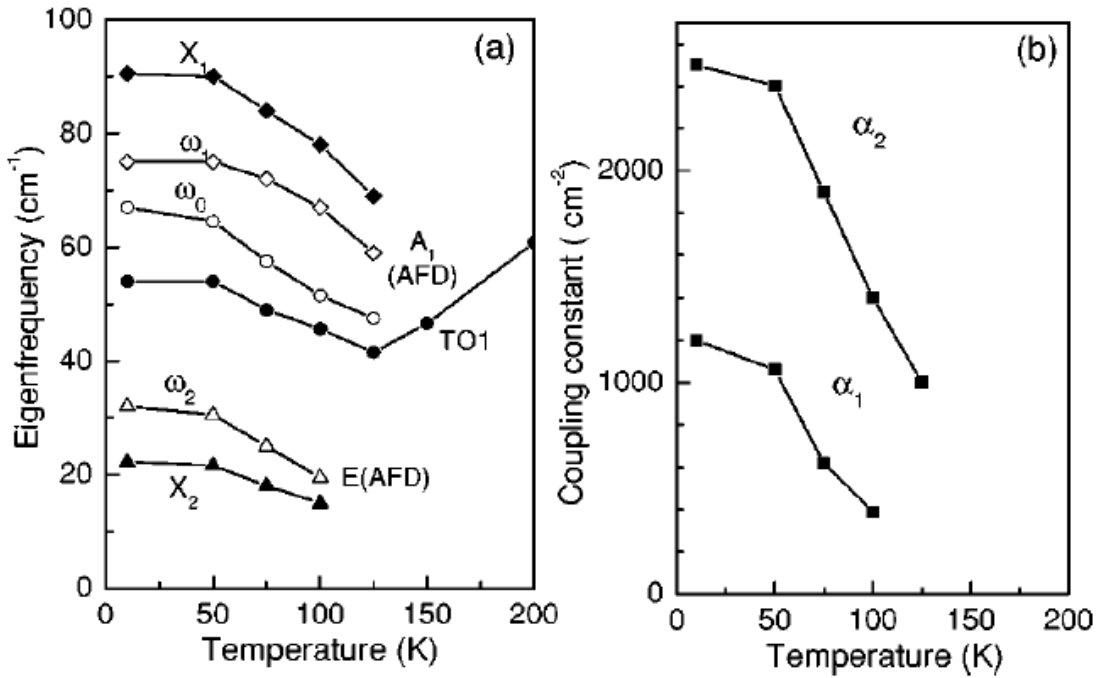

Fig. 4. Coupling between the TO1 mode and antiferrodistortive (AFD) $A_1$-E soft-mode doublet in the IR spectra of the quasi-epitaxial MOCVD thin film. The measured frequencies in increasing frequencies are $X_2$, TO1 and $X_1$, the bare (uncoupled) frequencies are $\omega_2$, $\omega_0$, $\omega_1$, respectively. $\alpha_1$ and $\alpha_2$ are real coupling constants between the TO1 and $A_1$ and E AFD mode, respectively. The in-plane ferroelectric transition appears near 120 K. (After [38]).

Raman scattering [50], because the field also activates the TO1 mode in the Raman response. Recently, the tunability on a thin film was studied up to the THz frequencies [51] and it was confirmed that the tunability does not depend on frequency and is fully due to the strong TO1 stiffening and appears up to room temperature (10 % permittivity decrease at 100 kV/cm).

Let us briefly discuss the soft mode behaviour in doped STO ceramics. A series of Bi-doped (up to 16.7%) samples substituting for Sr was studied by broad-band dielectric spectroscopy [52, 53, 54]. Strong stiffening of the TO1 mode in the whole temperature range was revealed using the time-domain THz and Fourier-transform IR spectroscopy (see Fig. 5). The effect was assigned to biasing the soft mode by the random $E$ field produced by the charged Bi ions, since the soft-mode eigenvector concerns predominantly the $TiO_6$ octahedra bending and does not involve the Sr(Bi) sublattice. Similar stiffening was observed and interpretation suggested in $KTaO_3$:Li crystals [55, 56, 57]. In addition, complex dielectric relaxation behaviour was observed and analyzed. Similarly, Mg and Mn doping on the dielectric spectra was extensively studied [58, 59]. Both ions can substitute for both Sr and Ti ions and the corresponding ceramics were prepared and studied separately. Substitution for Ti causes pronounced stiffening of the soft mode and decrease in permittivity, their temperature dependence and tunability. This can be understood as due to direct influencing the soft mode eigenvector. On the other hand, substitution for Sr (particularly by Mn [59]) introduces relaxation into the dielectric spectra due to a formation of polar clusters of off-centered Mn sites, similar to Bi-doping. Also the comparable soft mode stiffening was observed.

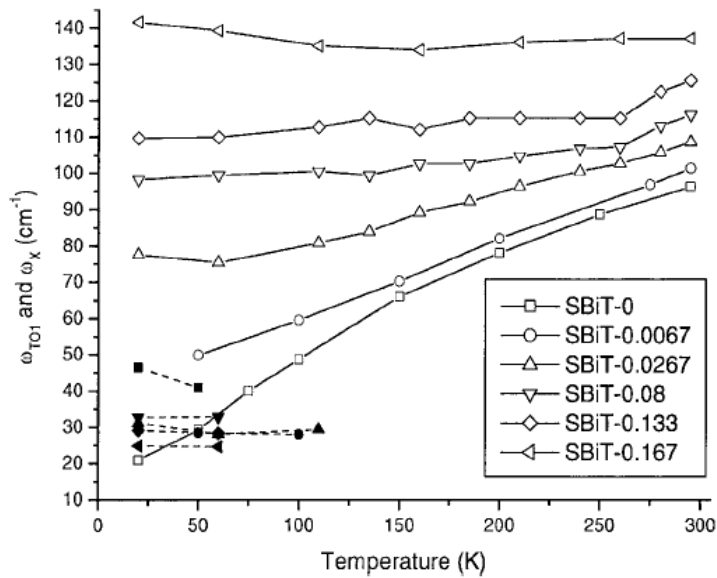

Fig. 5. Temperature dependences of the TO1 frequencies (open symbols) in STO:Bi ceramics. Full symbols denote the IR activated $E_g$-component of the AFD soft mode. (After [53]).

## 6. IR AND RAMAN SOFT MODE SPECTROSCOPY OF $BaTiO_3$ CERAMICS AND FILMS

Ferroelectric $BaTiO_3$ (BTO) ceramics were the first materials in which the grain-size effect on the dielectric response down to submicron size was carefully studied [60, 61, 62, 63]. In the ferroelectric phase, the size effect is dominated by the domain dynamics which varies with the grain size, whereas the paraelectric permittivity as well as the room-temperature TO phonon contribution is independent of size down to 1 μm [64]. However, for the submicron grain sizes a strong reduction of the permittivity, particularly around $T_C \cong 400$ K, was found. This effect was explained assuming a low-permittivity dead layer near the grain boundaries [61, 62, 63]. The brick-wall model yields good fits with dead-layer thicknesses 2-3 nm and permittivities 70-120 ($\varepsilon_{gb}/d \cong 40$ (nm$^{-1}$)), but no microscopic picture was suggested.

The dominant contribution to permittivity, as in STO, is provided by the TO1 soft mode, which is, however, overdamped [65, 66, 67, 68, 69]. It is well established since the early sixties that there is no appreciable dielectric dispersion in the paraelectric BTO crystals at least up to 60 GHz [66, 70, 71, 72]. Therefore, one can expect that the permittivity reduction in ceramics should be due to the effective soft mode stiffening, like in STO. However, due to the soft mode overdamping, this effect is much more difficult to reveal. This task was recently performed by studying carefully the temperature dependent IR reflectivity of 4 ceramics with the grain size from 10 000 down to 50 nm. [73]. The results, together with their fits, at room temperature and close above $T_C$ are shown in Fig. 6. The fits yield correct values for the paraelectric low-frequency permittivity. In the overdamped case, the soft mode frequency $\omega_{SM}$ is more accurately characterized by the frequency of maximum of the dielectric loss function, which is approximately given by $\omega_{SM}^2/\Gamma$ and is plotted for all the ceramics in Fig. 7. The frequencies obey Eq. 3 and in the paraelectric phase their temperature dependences agree with those of measured reciprocal permittivities. The size effect is shown up basically only in the down-shift of the extrapolated critical temperature (by more than 100 K in the case of nanoceramics), in agreement with the expectation from the brick-wall model [62]. Below $T_C$, which is not strongly grain-size dependent, the degenerate soft mode triplet strongly splits into $A_1$ and E components in the tetragonal and rhombohedral phase and into $A_1$, $B_1$ and $B_2$ components in the orthorhombic phase. The lowest component (E, respectively $B_2$) remains overdamped down to the orthorhombic – rhombohedral transition and is also shown in Fig. 7. All the phase transitions are still discernible down to 50 nm grain size, but some smearing and overlapping of phases is seen in the case of both nanoceramics, in agreement with the thermodynamic theory [74].

Micro-Raman measurements in back-scattering geometry on a similar set of dense ceramics with the grain size down to 50 nm were also recently published [75, 76]. The Raman data also confirmed the existence of all (smeared) phase transitions with some possible coexistence of phases. In Fig. 8 the reduced Raman spectra of 3 ceramics are compared with those of single crystal at selected temperatures in all the phases [76]. Tiny differences among the samples still await for their detailed discussion. Since the modes should loose their Raman activity in the paraelectric phase, it is difficult to trace their behaviour close to $T_C$. The striking fact that the broad peak near 270 cm$^{-1}$ ($A_1$ component of the soft mode!) remains observable in the paraelectric phase as a first-order Raman feature, was revealed in single crystals already in 1973 [77] and speaks for existence of polar clusters in the paraelectric phase. The same feature was revealed from detailed fits of the IR reflectivity [73]. It should be also noted that in the ferroelectric phases, only in case of nano-ceramics the response is averaged over several grains and therefore is independent of the measured spot. In case of larger grains and single crystals, the spectra are slightly dependent on the probed grain (domain) orientation.

Soft modes on sputtered epitaxial BTO films (and also mixed BST films up to 40 % of Sr) on MgO substrates were studied using polarized Raman scattering by Yuzyuk et al. [78]. Regardless of compressive strain, in the tetragonal ferroelectric phase both soft mode components followed well the single crystal behaviour, but the ferroelectric transition was smeared (by 10-20 K). On doping with Sr the overdamped E-component became underdamped and the huge $A_1$-E splitting (~240 cm$^{-1}$ at room temperature) decreased linearly with Sr concentration. Ostapchuk et al. [79, 80] also studied the thin films using both IR and Raman spectroscopy. One quasi-epitaxial and two polycrystalline films of different thicknesses on sapphire substrates were studied. No pronounced differences among the spectra were detected. The IR soft mode spectra show no complete softening at $T_C$, the loss maxima remain above 20 cm$^{-1}$. The smearing of the phase transitions and phase coexistence is even stronger than in nano-ceramics. The low-frequency component of the soft mode near 40 cm$^{-1}$ remains present down to 10 K, which indicates coexistence of the rhombohedral phase (where such a mode is shifted up to ~170 cm$^{-1}$) with the orthorhombic phase. The Raman spectra of polycrystalline films show presence of polar modes up to 150 K above the single-crystal $T_C$. Tenne et al. [81] studied epitaxial PLD films on STO substrates with SrRuO$_3$ electrodes by polarized Raman scattering and compared them with single crystals. They concluded that due to the tensile strains from the electrode, in the measured temperature interval 5-325 K the film exhibits a single ferroelectric phase with the orthorhombic structure and in-plane polarization. They revealed also an overdamped soft mode component below 50 cm$^{-1}$ down to 5 K without appreciable temperature changes.

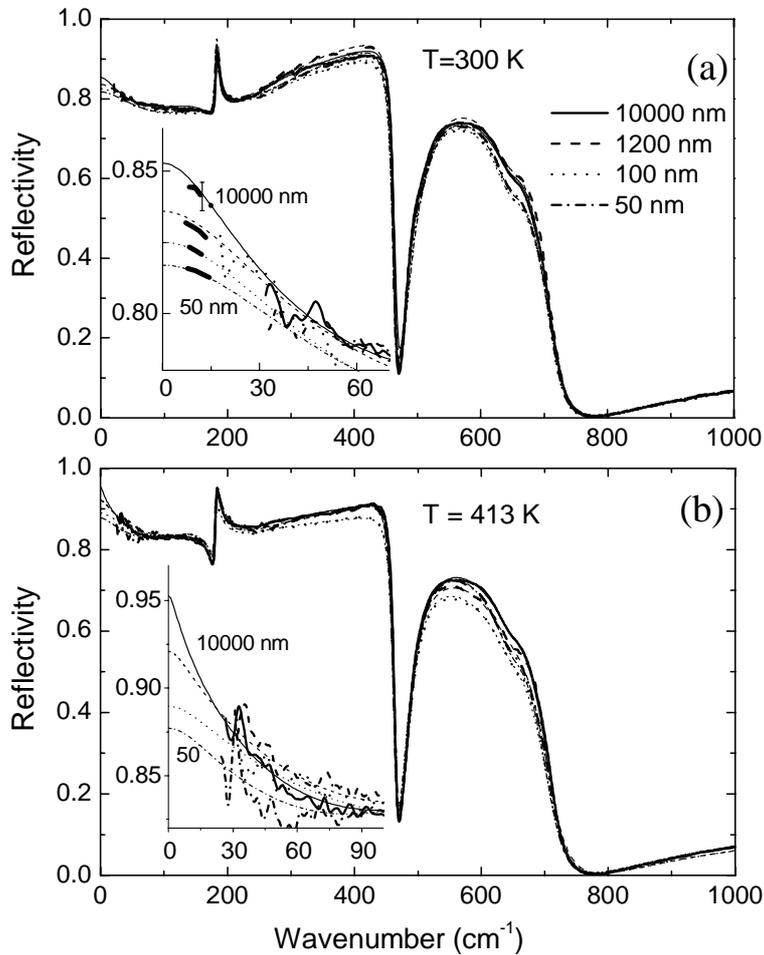

Fig. 6. IR reflectivity spectra of BTO ceramics with different grain size: 50, 100, 1200 and 10000 nm (a) in the tetragonal phase at T = 300 K and (b) in the cubic phase at 413 K. Thick lines indicate experimental data, thin ones the fitting curves. Full, dashed, dotted and dashed-dotted lines correspond to 10000, 1200, 100 and 50 nm grain size, respectively. Thick short lines in the inset of part (a) represent the THz data. (After [73]).

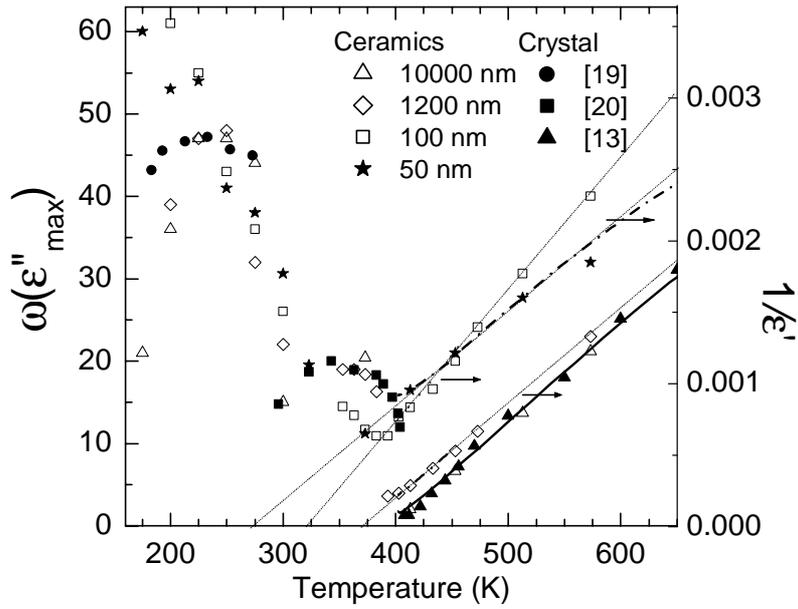

Fig. 7. Frequencies of the soft-mode loss maxima (left-hand-side axis) and inverse permittivity at 1 MHz denoted by lines (right-hand-side axis) as the function of temperature. Different symbols mark the results obtained on ceramics with different grain size or on various single crystals. Dotted lines indicate fits to the Curie-Weiss law. (After [73]).

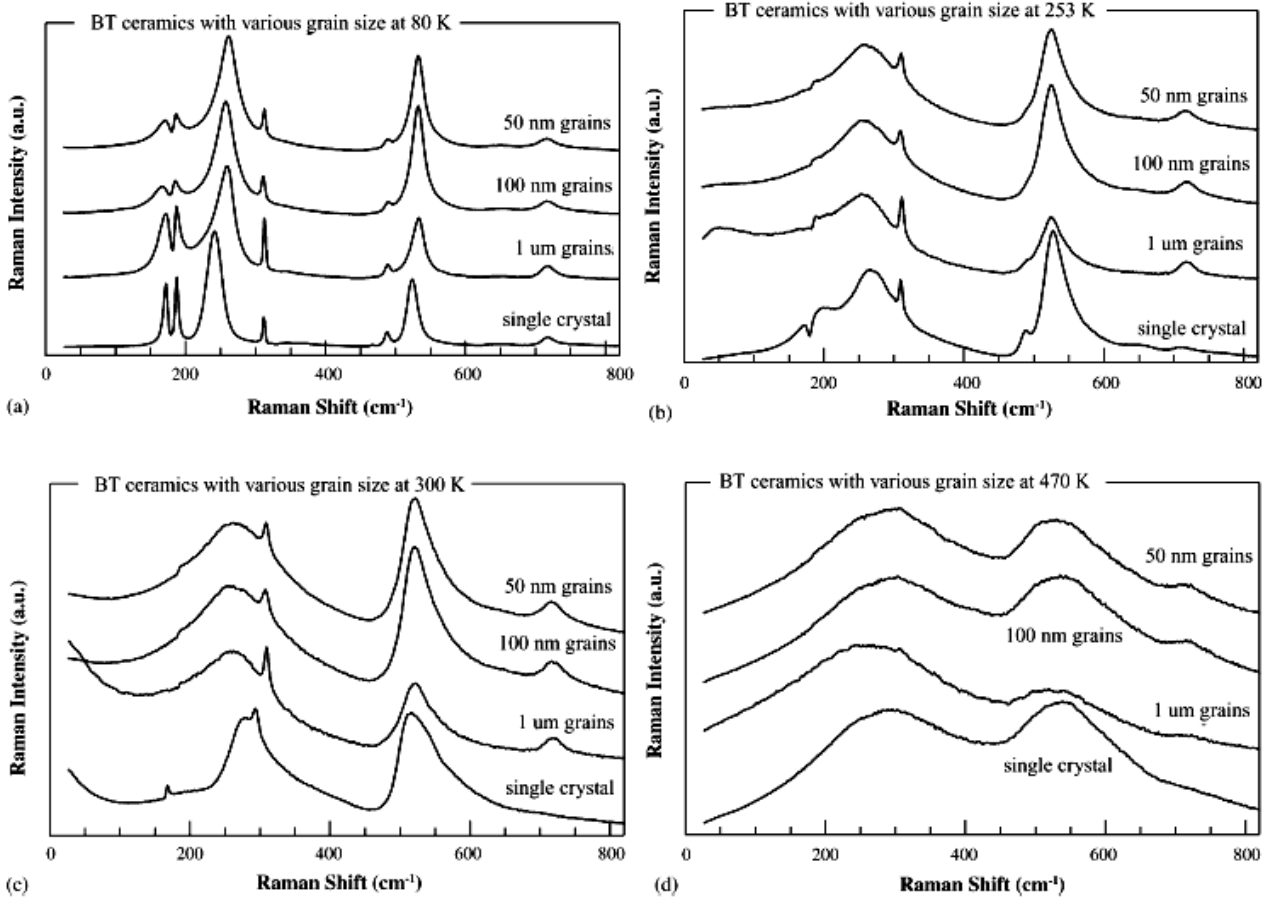

Fig. 8. Micro-Raman spectra of BTO ceramics with various grain sizes compared to those of single crystal at selected temperatures in all the phases. (After [76]).

## 7. BROAD-BAND DIELECTRIC, INFRARED, RAMAN AND NEUTRON SPECTROSCOPY OF PMN RELAXOR FERROELECTRICS

$PbMg_{1/3}Nb_{2/3}O_3$ (PMN) has been intensively studied during half of the century as a model of relaxor ferroelectrics (earlier called ferroelectrics with diffuse phase transitions) and the interest in this materials was still substantially enhanced after publishing the paper by Park and Shrout [82] who discovered ultrahigh strain and piezoelectric response in $PbMg_{1/3}Nb_{2/3}O_3$-$PbTiO_3$ (PMN-PT) and $PbZn_{1/3}Nb_{2/3}O_3$-$PbTiO_3$ (PZN-PT) single crystals.

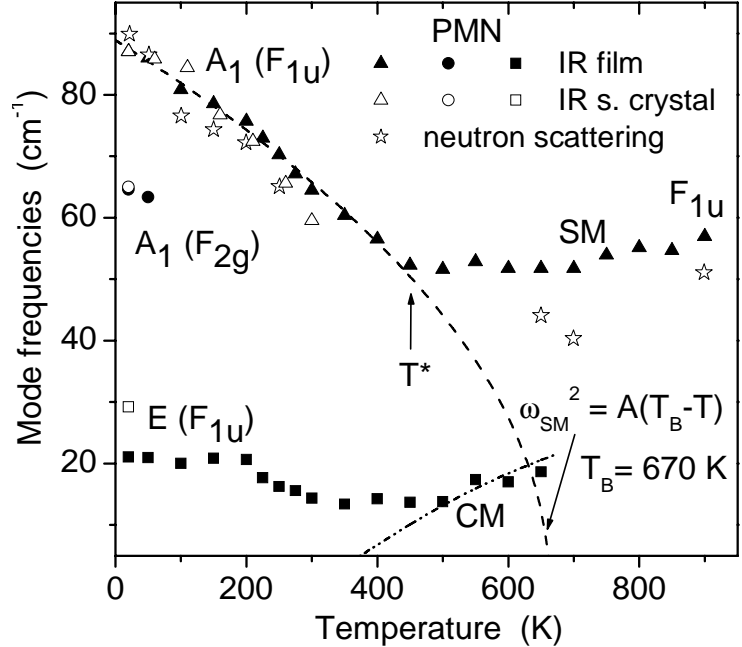

Fig. 9. Temperature dependences of the polar mode frequencies observed in IR spectra of PMN thin film [84] (solid points), single crystal [29, 83] (open points) and in inelastic neutron scattering of single crystal [85] (stars). SM and CM are abbreviations for soft and central modes, respectively. Dashed line shows the Cochran fit to the temperature dependence of the $A_1$ soft-mode component with $A = 11.9$ $K^{-1}$ and $T_B$=670 K.

Result of extensive IR studies of polar phonons in PMN single crystal [29, 33, 83], thin film [84] as well as data from inelastic neutron scattering (INS) studies [85] are summarized in Fig. 9. In the film, the polar-mode frequencies are not appreciably influenced by the grain boundaries and possible strains and defects, and the phonon frequencies in the PMN film do not differ appreciably from those in single crystal. The situation differs from the situation in incipient ferroelectrics [38] where the soft mode in films is strongly influenced by the strain and microstructure.

In INS spectra, the soft mode is not resolved below the Burns temperature ≈ 600 K down to the freezing temperature $T_f$ ≈ 200 K [85]. This difficulty is related to the so called phonon waterfall effect, first described in PZN-PT relaxor ferroelectric [86]. The INS spectra in this case suggest a sharp drop (waterfall) of the soft optic branch dispersion curve into the lower acoustic branch at a finite phonon wavevector $q_{wf}$ ≈ 0.2 Å$^{-1}$. The value of $q_{wf}$ was correlated with the size of the polar nanoclusters [85, 86], but later it was revealed [87] that $q_{wf}$ is different in different Brillouin zones so that its value can hardly provide a measure for the size of polar nanoclusters. The waterfall effect was later explained by coupling of acoustic and heavily damped optic phonon modes [87]. In this model, the value of $q_{wf}$ depends on the measured Brillouin zone, because the INS dynamical structure factor of bare acoustic branch strongly differs in different Brillouin zones.

One can see in Fig. 9 that the soft TO1 mode of $F_{1u}$ symmetry splits into two components below $T_B$: $A_1$ component hardens on cooling following roughly the Cochran law with the extrapolated critical temperature near $T_B$, while the heavily damped E component remains very soft (below 1 THz) and nearly temperature independent down to helium temperatures. The reason for splitting of the soft mode is the strong uniaxial anisotropy of the polar clusters [33]. The $A_1$ mode softening is not full, it levels off above $T^* \cong 450$ K. Anomalies near $T^*$ were observed also in structural studies [88], Raman scattering [89] as well as in acoustic emission studies [90]. The interpretation of anomalies near $T^*$ is still under discussion, but the authors mostly agree that it is connected with enhanced growth and stronger interaction among the polar clusters on cooling.

$A_1$ mode near 65 cm$^{-1}$ at low temperatures probably stems from the $F_{2g}$-symmetry mode, which is Raman active even in the cubic phase due to the chemical clusters which locally double the unit cell [89]. This mode is resolved in the IR spectra only below 100 K, when the damping of neighbouring modes becomes lower. Dotted line in Fig. 9 marked by CM shows schematically how the additional relaxation central mode shifts from the MW region into the THz region and overlaps with the E soft-mode component and finally disappears from the spectra above $T_B$ (see the discussion below). Detailed discussion of the full IR spectra is beyond the scope of this article, nevertheless the IR spectra of many complex perovskites have been recently reviewed in [34].

Lattice dynamics of PMN was also intensively studied by means of Raman scattering (see for example [91, 92, 93]). As the simple cubic $Pm\overline{3}m$ perovskite structure allows no Raman active modes, all rather intensive peaks observed in Raman spectra are indications of at least local deviations from this ideal average structure. The most obvious source of such structural perturbations is the short-range 1:1 chemical order of $Mg^{2+}$ and $Nb^{5+}$ cations at the perovskite B-sites, but local polar distortions can activate the Raman modes, too [92]. High temperature Raman spectra are usually interpreted as stemming from the 1:1 ordered chemical clusters having the $Fm\overline{3}m$ structure with doubled unit cell. Additional Raman peaks appearing at low temperatures could be associated with polar clusters. If we assume that the clusters have rhombohedral R3m (Z=2) symmetry (still not experimentally confirmed), then 16 vibration modes both Raman and IR active are expected [94]. Certain indications of the expected Raman activity of IR modes have been indeed noticed at low temperatures, but the Raman intensity of such modes is rather small. The soft mode was better resolved in Raman spectra of electrically poled PMN-xPT single crystal [95]. Raman scattering spectroscopy was also successfully used for the study of high-pressure phase transitions in PMN [96, 97], but their discussion is beyond the scope of this review.

Near $T_{max} \cong 240$ K PMN shows a huge low-frequency dielectric anomaly with $\varepsilon'_{max} \cong 18000$ [29], but one can see in Fig. 9 that there is no anomaly of the optic soft mode near 240 K. Dielectric contribution of all polar phonons is less than 300 [84], therefore the dielectric anomaly near $T_{max}$ should be caused by a dielectric dispersion below the phonon frequencies which we call central mode. Complex dielectric spectra of PMN were investigated by many authors (see [98] and references therein). Coverage of the wide frequency interval from 10$^{-3}$ up to polar phonon frequencies has been published only recently [98] – see Fig. 10. Above room temperature, the dispersion appears mainly in the microwave range. On cooling it slows down, broadens and finally splits into two overlapping components. The first one, probably caused by flipping of polar clusters, slows down into the sub-Hertz range near the freezing temperature ($T_f \cong 200$ K), but surprisingly remains present in the dielectric spectra even below $T_f$. Second part, interpreted as due to breathing of polar clusters, anomalously broadens below $T_f$ giving rise to frequency independent losses from Hz up to at least GHz range. Such constant losses are typical for all relaxors and remain present in the spectra down to liquid He temperatures decreasing exponentially on cooling [29]. Frequency independent losses are equivalent to broad uniform distribution function of relaxation frequencies, which probably appears due to the influence of random fields (caused by chemical disorder in the perovskite B-sites as well as polar clusters) on the distribution of activation energies for hopping of disordered Pb cations.

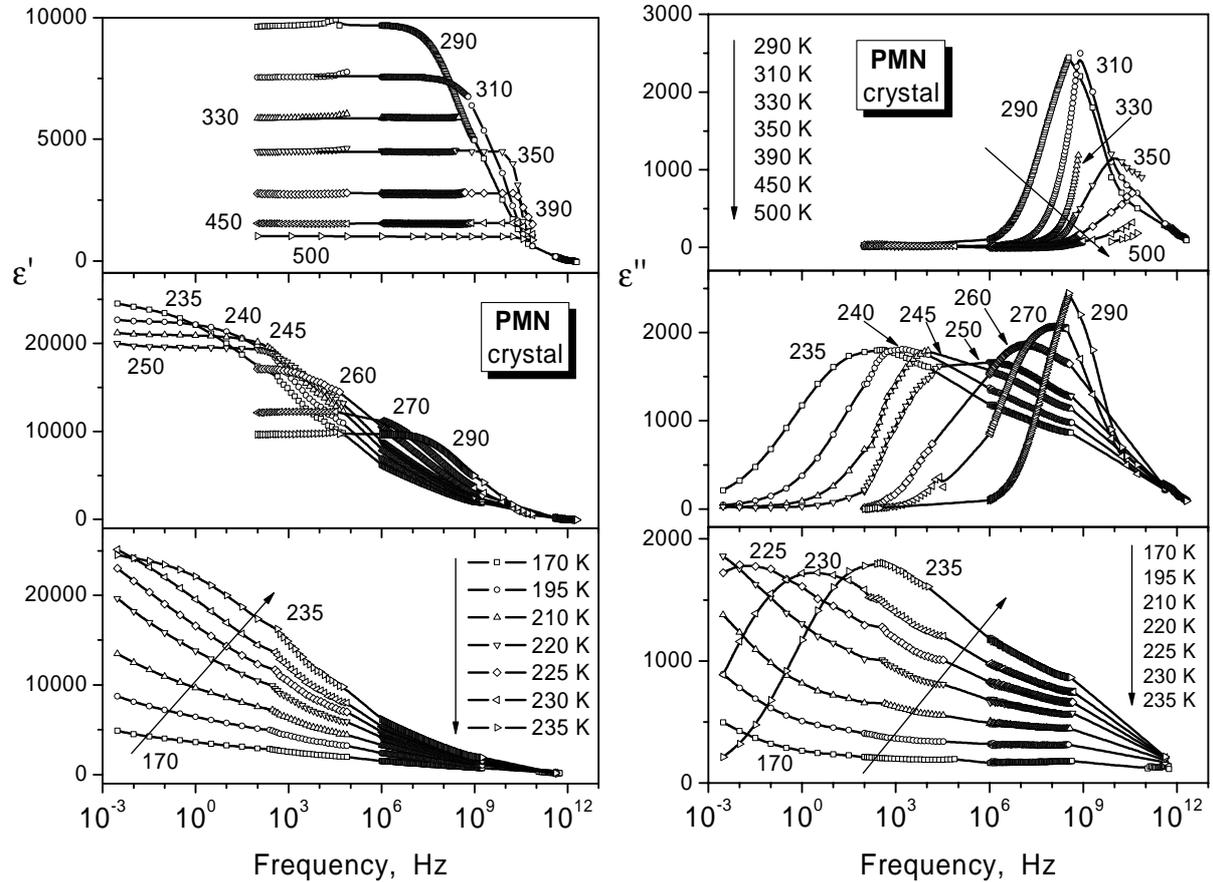

Fig. 10. Frequency dependences of the dielectric permittivity $\varepsilon'$ and dielectric loss $\varepsilon''$ of PMN single crystal at various temperatures. (After [98]).

PMN ceramics with coarse grains show qualitatively the same dielectric relaxations as single crystals, but the value of permittivity maximum $\varepsilon'_{max}$ is by 10-20% lower [99]. If the grain size of the ceramics reduces from 6 to 0.3 µm (see Fig. 11), $\varepsilon'_{max}$ dramatically decreases by about one order of magnitude [100]. This decrease was also successfully explained using the series capacity (essentially brick-wall) model with dead layers, consisting of PbO grain boundaries few nm thick and permittivity of ~20 [100]. In nano-grain PMN ceramics $\varepsilon'_{max}$ decreases even more and the relaxor behaviour totally disappears below ~30 nm grain size, $\varepsilon'_{max}$ reaching values only of ~115 for the ~15 nm grain size [101]. In Ref. [101] this effect was explained by grain-size dependent random stresses and electric fields which destabilize the polar clusters. Actually, it was shown by TEM that the size of polar nanoclusters decreases with diminishing grain size in PMN nanoceramics [102], therefore some stiffening of the polar clusters and reduced contribution into the bulk permittivity might be also expected.

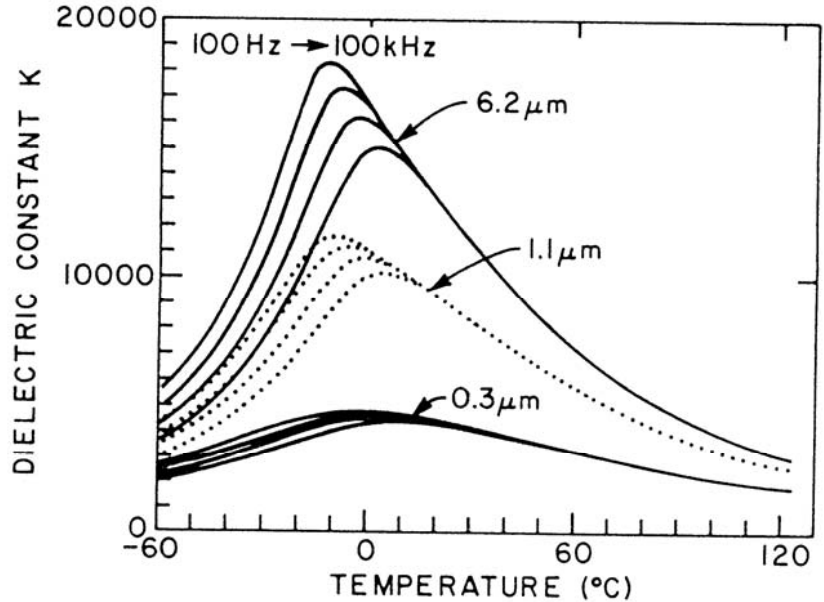

Fig. 11. Temperature dependence of the permittivity of PMN ceramics with various grain sizes. (After [100]).

PMN thin films exhibit also considerably lower $\varepsilon'_{max}$ than the single crystals. For example, $\varepsilon'_{max}$ in 500 nm thick film deposited on sapphire substrate is one order of magnitude lower than in the crystals and temperature of $\varepsilon'_{max}$ is also shifted by 30 K up in the thin film [84]. It is still not clear if it is due to the intrinsic size effect, grain boundary contribution, interface layer or stress (strain) between the thin film and substrate. Probably all these mechanisms contribute to the permittivity reduction. Nevertheless, it was established that the polar phonon contributions to permittivity are not influenced in the films [84]. It means that the whole change in the effective permittivity of the film is due to the change of the dielectric relaxation strength, as a consequence of the change in the distribution function of relaxation frequencies and probably simultaneous stiffening of the mean relaxation frequency. All these features are in agreement with the expectations from the brick-wall model for core-shell composites applied to the ac response, since the dielectric strength of relaxations is much larger than the phonon contribution and therefore practically only the former one is influenced by the depolarization field. So we could suggest that the dead layers play the dominant role in the reduction of the dielectric response even in relaxor ferroelectric ceramics and films.

## 8. DIELECTRIC AND IR SPECTROSCOPY OF ORDERED AND DISORDERED PST CERAMICS AND FILMS

Lead scandium tantalate $PbSc_{1/2}Ta_{1/2}O_3$ (PST) is also a model representative of relaxor ferroelectrics, but it exhibits several differences compared to PMN. Unlike PMN, PST undergoes spontaneous ferroelectric phase transition on cooling without bias electric field. Its temperature $T_c$ depends on the degree of ordering of the B-site ions ($Sc^{3+}$ and $Ta^{5+}$), which can be controlled by sample annealing. Disordered PST undergoes ferroelectric transition near 270 K, while $T_c$ of the ordered sample appears near 300 K without typical relaxor behaviour above $T_c$ (see Fig. 12) [103, 104]. The value of $\varepsilon'_{max}$ strongly depends on the degree of B-site order [103, 104, 105], therefore the grain size dependence of $\varepsilon'_{max}$ was not investigated.

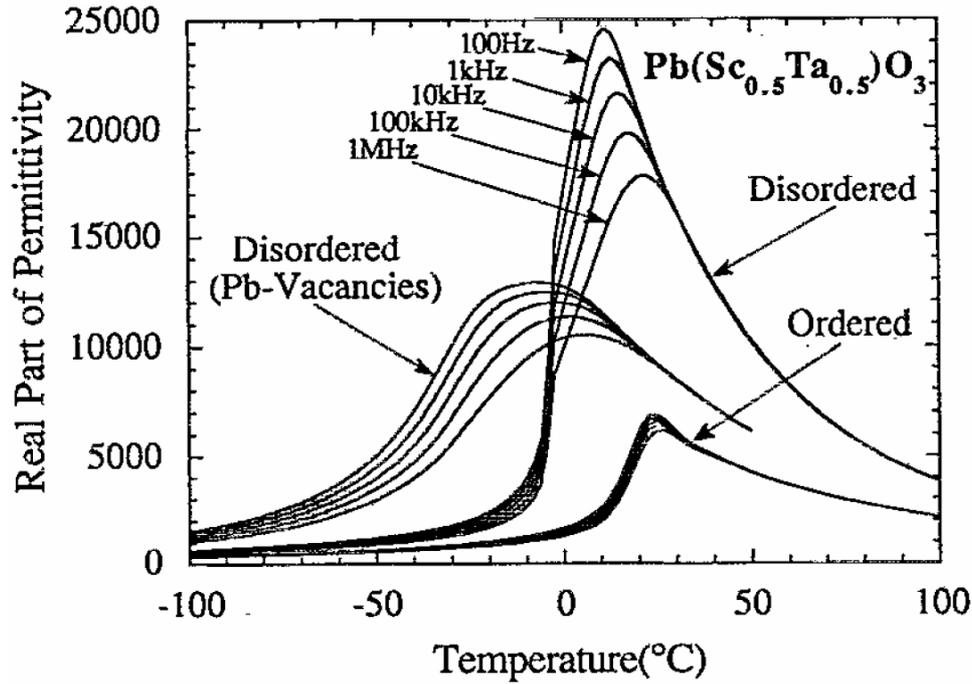

Fig. 12. Temperature dependence of the permittivity of PST ceramics with different order in B sites and with Pb vacancies. (After [104]).

Polar phonons were investigated in PST ceramics [106] and thin films [107] by means of IR spectroscopy. Low-frequency phonon frequencies in ceramics and thin films do not appreciably differ, but the accuracy of FIR transmission of thin films is higher than that of FIR reflectivity of ceramics, therefore we will discuss here in more detail the results obtained from the films. Fig. 13 shows temperature dependence of the low-frequency polar modes below 90 cm$^{-1}$ in the 78 % ordered sample. Both $A_1$ modes soften on heating and $A_1(F_{1u})$ apparently disappears from the spectra above $T_c$. Nevertheless, it overlaps with $A_1(F_{2g})$ mode, whose intensity reduces on heating and finally disappears from the IR spectra above $T_B$ (but it remains in the Raman spectra [108]). $A_1(F_{1u})$ mode shows minimum frequency at 550 K and than starts to slightly harden on further heating and attains the $F_{1u}$ symmetry above $T_B \approx 700$ K. Temperature 550 K probably corresponds to $T^*$ temperature, known already from PMN, below which the polar clusters abruptly increase. Central (relaxational) mode (due to dynamics of polar clusters) and the E component of the $F_{1u}$ soft mode appear (overlapped) below $T_B$ and the central mode softens on cooling. Below $T_C$ the relaxation slows down below the FIR range and only the highly damped E mode remains in the FIR spectra.

In Ref. [107], four different thin films were investigated with various B-site order (from completely disordered to 78 % ordered film). It was shown that the degree of cation order has a large influence on the damping of all phonons, but the temperature dependences of both components of the soft mode was almost the same. It seems that the Burns temperature $T_B$ (temperature where $F_{1u}$ mode splits into $A_1$ and E component) is not dependent on the degree of the B-site order, however our very recent results [109] show that $T_B$ can depend on the film substrate.

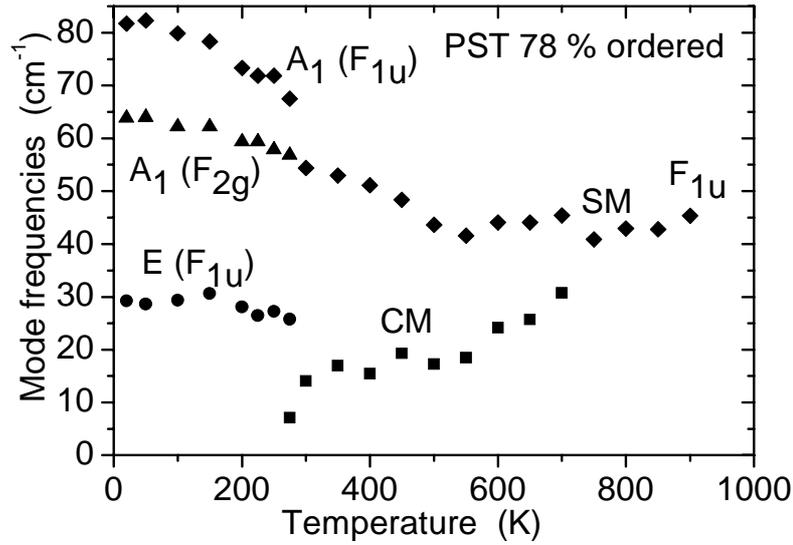

Fig. 13. Temperature dependences of the polar mode frequencies below 90 cm$^{-1}$ in the 78% ordered PST thin film. SM and CM mark soft mode and central modes, respectively. (After [107]).

It is worth to note that the $A_1(F_{1u})$ mode softens towards the Burns temperature and not towards the ferroelectric phase transition near 300 K. It shows that the ferroelectric phase transition is not of a displacive type and the main precursor fluctuations of this phase transition may be of relaxational type. As a matter of fact, broad-band dielectric studies of PST ceramics performed up to 33 GHz and combined with submillimetre and IR measurements revealed dielectric relaxation with two well-separated components [106, 110, 111]. Frequency of the first component remains in the THz range and is almost temperature independent (presumably the E component of the soft mode), while frequency of the second component slows from the THz range [107] down to $3.10^8$ Hz at 305 K [111]. Below $T_C$ it remains in the $10^8 – 10^9$ Hz range, while the dielectric strength strongly decreases and the spectrum anomalously broadens on cooling [111]. This latter component is obviously associated with the dynamics of polar clusters above $T_C$, which should abruptly grow into the ferroelectric domains below $T_C$. Nevertheless, the typical order-disorder increase of its relaxational frequency below $T_C$ is not observed here. The nonzero relaxation strength below $T_C$ could be understood by a small amount of polar clusters, which remain among the ferroelectric domains and gradually disappear on cooling. Such situation is analogous to that occurring in PMN-PT system, where the relaxor effects are seen also mainly above $T_C$ and the polar clusters also seem to coexist with the ferroelectric domains below $T_C$ [112].

PST thin films exhibit an order of magnitude lower permittivity as compared to bulk ceramics. This has often been attributed to passive interfacial layers in series with the film capacitance. However, the recent study of the out-of-plane and in-plane dielectric response revealed no differences between them, which means that the former reason is not valid and the origin of lower permittivity should be in the structural and microstructural properties of the relaxor film itself [113].

# 9. SOFT MODE BEHAVIOUR IN OTHER FERROELECTRICS AND RELATED MATERIALS

Similar soft mode behaviour as in PMN and PST has been recently observed also in the following perovskite relaxor ceramics and thin films: $PbMg_{1/3}Ta_{2/3}O_3$, $PbSc_{1/2}Nb_{1/2}O_3$, PMN-PT and PLZT [109]. In all these materials, splitting of the $F_{1u}$ soft mode below $T_B$ was revealed, hardening of its $A_1$ component on cooling, activation of the Raman active $F_{2g}$ mode in the low-temperature IR spectra near 60 cm$^{-1}$ and appearance of the dielectric relaxation below $T_B$ in the THz range, overlapped with the E-component of the soft mode at high temperatures, but its slowing down and broadening on cooling. At low temperatures, the broadening exceeds the measured frequency range giving rise to frequency independent losses up to the GHz range. It is important to stress that the soft mode does not soften to $T_C$ or $T_{max}$, but approximately to $T_B$, which indicates that it acts as the local ferroelectric soft mode inside the polar clusters. On the other hand, the soft mode never softens completely, but it levels off above $T^*$. Moreover, the $A_1$ component of the soft mode always reduces its effective IR damping on heating above $T^*$, although the phonon damping usually increases with the rising temperature. This indicates some inhomogeneity in the sample (inhomogeneous phonon broadening) due to the polar clusters which decreases on heating towards $T_B$.

Soft-mode behaviour in many other ferroelectrics and antiferroelectrics was described in detail in the review [10]. In addition to relaxor ferroelectrics, lattice dynamic properties of $PbZrO_3$, $AgNbO_3$, $LiNaGe_4O_9$, $LaBGeO_5$, $Cd_2Nb_2O_7$, $SrBi_2Ta_2O_9$, $Sr_2Nb_2O_7$, $Ba_2NaNb_5O_{15}$, betaine calcium chloride dihydrate, $BiScO_3$-$PbTiO_3$ and $Rb_{1/2}(ND_4)_{1/2}D_2PO_4$ were discussed. Some of the more recent results on relaxors, particularly on $Na_{1/2}Bi_{1/2}TiO_3$, $Sr_xBa_{1/x}Nb_2O_6$ and $Ba_2LnTi_2Nb_3O_{15}$ were reviewed in [34]. Some of the results were obtained on ceramics or on both ceramics and single crystals ($PbZrO_3$, $Na_{1/2}Bi_{1/2}TiO_3$, $SrBi_2Ta_2O_9$ $BiScO_3$-$PbTiO_3$, $Ba_2LnTiNb_3O_{15}$), but no detailed comparison of the soft mode behaviour with that in single crystals was provided, with the exception of $PbZrO_3$ (see next paragraph). One can summarize the soft mode properties as follows: No pure displacive phase transition, where the dielectric anomaly near $T_C$ would be caused only by phonon softening, was observed. In all cases only more or less partial phonon softening was observed and some additional dielectric dispersion (either directly observed or needed to explain the low-frequency dielectric data) appears below the polar phonon range, typically in the $10^{10}$-$10^{11}$ range.

Finally, let us mention the very recent THz spectroscopy results obtained on high-temperature Bi-layered ferroelectrics and relaxors with Aurivillius structure (first obtained on $SrBi_2Ta_2O_9$ [114], and then on $Bi_4Ti_3O_{12}$, $SrBi_2Nb_2O_9$, $BaBi_2Nb_2O_9$, and $Sr_{0.5}Ba_{0.5}Bi_2Ta_2O_9$) [115]. The measurements were performed with single crystals as well as with ceramics (latter two materials), but no special differences were mentioned. In all the compounds very low-frequency and low-damped polar modes were found which gradually soften and broaden on heating up to 900 K, but they did not show any anomaly near the ferroelectric transitions. This was unexpected, particularly for bismuth titanate $Bi_4Ti_3O_{12}$, in which no disorder was observed in structural studies. The dielectric anomaly at $T_C$ or $T_{max}$ cannot be ascribed to the lattice, which contributes by a few hundreds to the permittivity at most. We assume again that polar clusters are responsible for the additional central mode dispersion, directly revealed in the THz range in the case of $SrBi_2Ta_2O_9$ [114] and in the GHz range in the case of relaxor $BaBi_2Nb_2O_9$ [115]. As in PMN-PT and PST, the origin of the ferroelectric transition is a sudden increase in the dynamic polar cluster size changing into quasistatic ferroelectric domains rather than a classical order–disorder mechanism.

# 10. EFFECT OF CRACKS ON THE DIELECTRIC RESPONSE OF PbZrO$_3$ CERAMICS

PbZrO$_3$ is a classic material with a single antiferroelectric transition at 508 K. The critical antiferroelectric soft mode is not know, but the (noncritical, secondary) ferroelectric softening was studied on single crystals, thin films [116] as well as ceramics (PZ1) [117]. No substantial differences among the samples were mentioned. The temperature evolution of the dielectric function obtained from the FIR reflectivity on ceramics PZ1 is shown in Fig. 14. Note again combination of partial mode softening and appearance of the central mode in the $10^{11}$ Hz range, which is responsible for the substantial part of the dielectric anomaly.

Later on, another single-phase ceramics, PZ2, processed by hot pressing, was studied [118]. The density of both ceramics was more than 98 %, but the microstructure was different. PZ1 displayed small and very irregular grains, whereas PZ2 showed large (over 10 μm) regular grains, but with possible nanocracks along some of the grain boundaries. The dielectric properties of both ceramics differ dramatically (see Fig. 15), particularly near the dielectric maximum at $T_C$. The dispersion up to 1 GHz is very weak in both ceramics and the transition temperatures do not differ from each other. It was possible to reproduce the data of both ceramics using the single crystal dielectric function at 510 K (close above $T_C$) and a simple brick-wall model with very small crack-type porosity of 0.02 and 0.29 % for PZ1 and PZ2, respectively (see Fig. 15). Let us note that the reason for the smaller PZ2 permittivity due to the cracks is basically stiffening of the relaxational central mode (by a factor of 4 which reduces the permittivity by about the same factor). This pronounced effect on the dielectric response asks for some caution in the dielectric studies of ceramics.

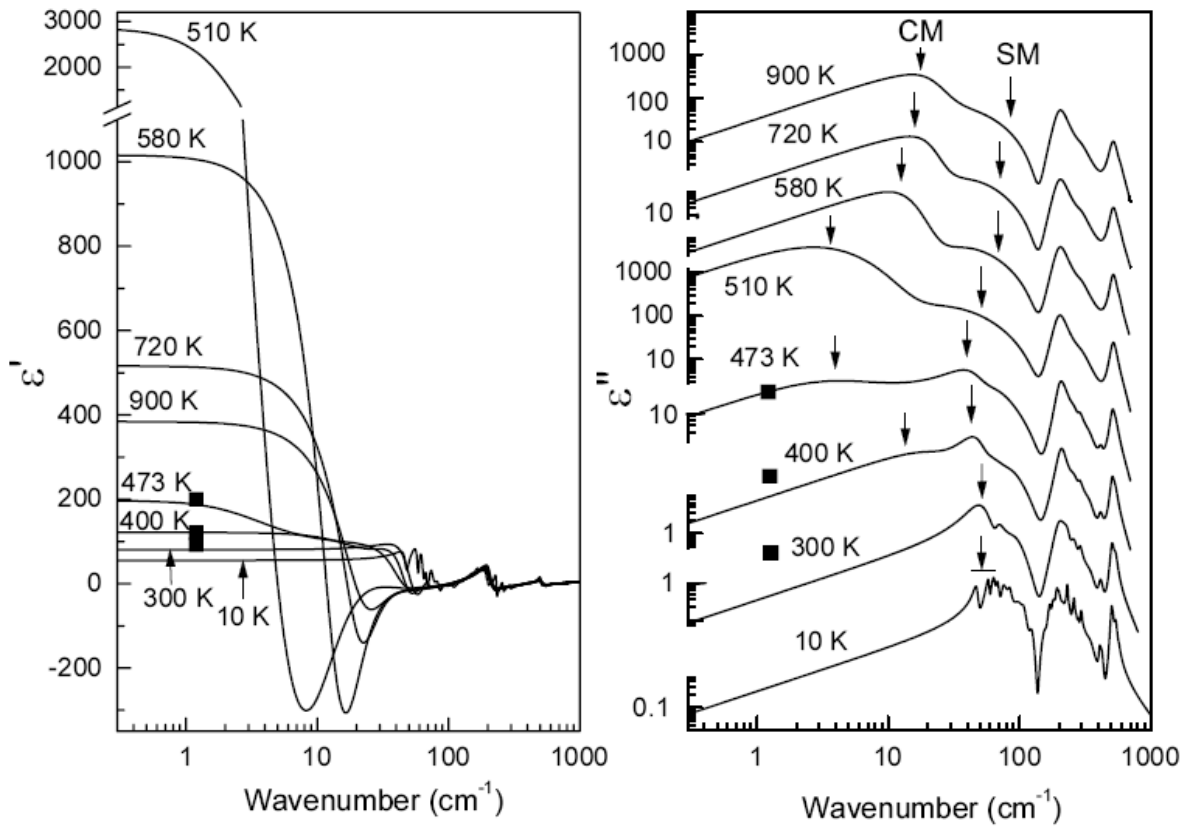

Fig. 14. Dielectric function obtained from the fit to IR reflectivity of PbZrO$_3$ ceramics. Full squares indicate the microwave data at 36 GHz. (After [117]).

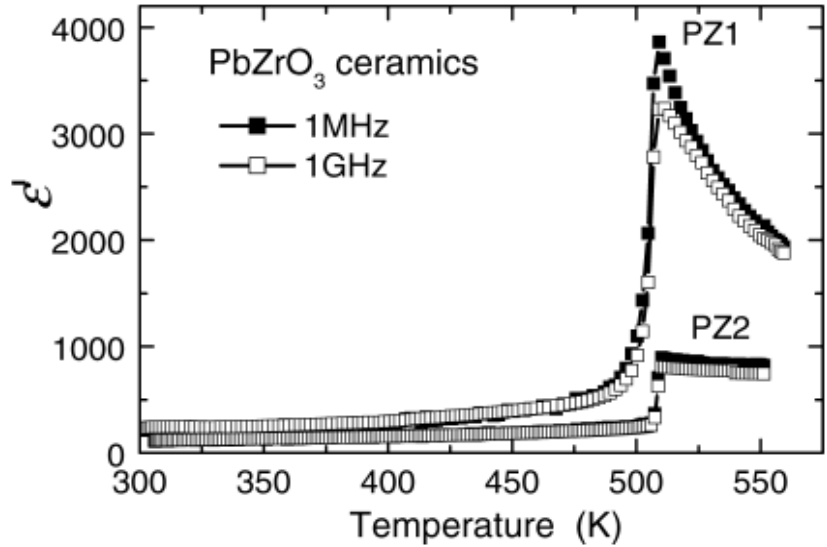

Fig. 15. Temperature dependence of the permittivity of PZ1 and PZ2 ceramics. The substantially smaller permittivity of PZ2 is due to the presence of nanocracks. (After [118]).

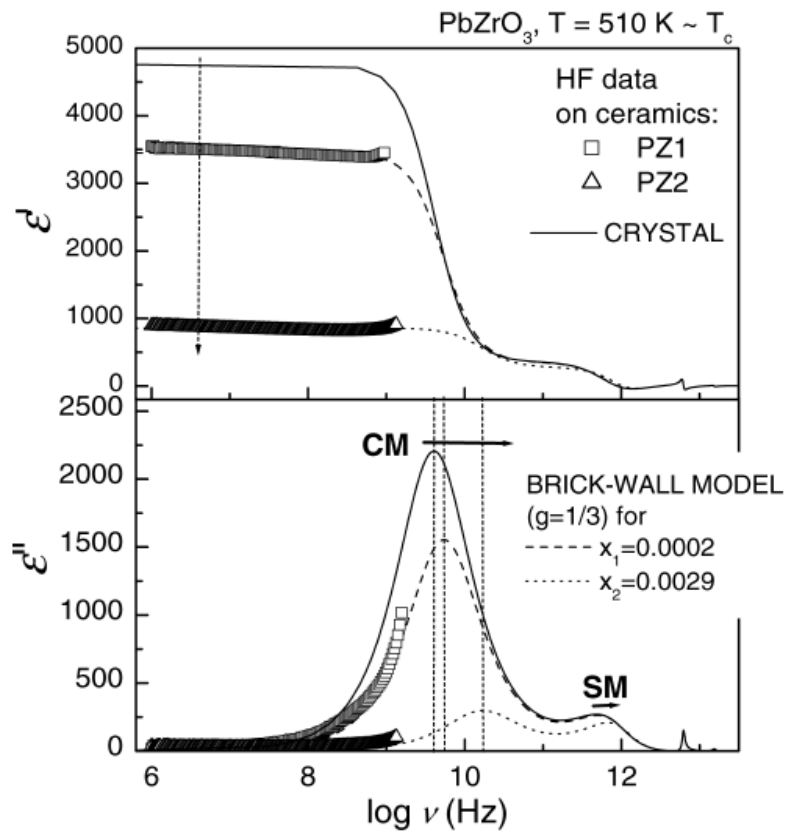

Fig. 16. Dielectric function of PZ1 and PZ2 ceramics compared to that of single crystal. Solid line is obtained from the fit to IR reflectivity of $PbZrO_3$ single crystal, whereas the dashed and dotted curves are calculated using the crystal data and brick-wall model with $x_1$ and $x_2$ volume fraction of the crack-type porosity of the PZ1 and PZ2, respectively. Vertical dotted lines denote the central mode frequencies (maximum in the dielectric loss spectra) for the three samples. (After [118]).

## 11. SUMMARY AND CONCLUSIONS

IR and Raman spectroscopic data on ferroelectric and related ceramics and thin films bring important information about the soft polar phonon mode behaviour in these materials. This data allow us to discuss the dynamic origin of the phase transition. It was revealed that in pure incipient ferroelectric STO (not reaching the ferroelectric transition down to 0 K), the soft mode fully accounts for the temperature dependent dielectric response as well as for its tuning in electric field, which is valid for ceramics, probably films as well as for single crystals. Such a crystal structure is only weakly anharmonic and the approaching ferroelectricity is purely displacive. Also in BTO the THz soft mode fully accounts for the dielectric anomaly in the paraelectric phase in ceramics as well as in crystal, but it is overdamped so that the experimental accuracy is lower than in STO. In most other cases, the ferroelectric transition was only partially (or not at all) driven by the complete phonon mode softening. Additional dielectric dispersion below the polar phonon range (central mode) accounts for the main part of the dielectric anomaly near $T_C$. This is also true for ceramics as well as for single crystals and can be assigned to crossover from displacive behaviour (far from $T_C$ dominated by the soft phonon contribution to the dielectric response) to order-disorder regime (close to $T_C$ dominated by the central mode contribution). The central mode can be assigned to dynamic fluctuations of polar clusters, appearing in some temperature range near $T_C$ or below the Burns temperature in the case of relaxor ferroelectrics. Cases where the soft mode shows no anomaly at $T_C$ at all (e.g. relaxors with a spontaneous ferroelectric transition or Bi-layered ferroelectrics with Aurivillius structure) represent a new category, where the phase transition is connected merely with a sudden increase (percolation?) in the polar cluster size. Similar feature appears in relaxor ferroelectrics, in which the permittivity maximum (smeared and frequency dependent) is not revealed in any soft mode anomaly.

Important issue concerns the granularity of ceramics and polycrystalline films, which usually produces a pronounced dielectric inhomogeneity in high-permittivity materials. It can be treated using effective medium approximation for core-shell composites. The generally observed reduction in the dielectric response on decreasing grain size can be well described by thin dead grain-boundary layers with small (temperature independent) permittivity. Particular caution must be paid to possible nano-cracks along the grain boundaries, which dramatically reduce the effective response. The microscopic nature of dead layers was studied only in the case of STO ceramics. It appears that it is connected with a strong oxygen deficit in the grain-boundaries which creates a polarized region in the close vicinity. The corresponding brick double-wall model can well explain all the observed dielectric and spectroscopic features.

The reduction of the dielectric permittivity with decreasing grain size has its dynamic counterpart in appreciable stiffening of the strongest effective mode frequency, revealed in the dielectric spectra. Such effective mode frequencies may differ from those observed in the Raman response, because this technique probes much higher phonon wavevectors. The corresponding phonon wavelengths might be comparable or smaller than the grain size, which goes beyond the applicability of the effective medium approach.

**ACKNOWLEDGEMENTS**
The work was supported by the Grant Agency of the Czech Republic (Projects Nos. 202/04/0993 and 202/06/0403).